\PassOptionsToPackage{unicode}{hyperref}
\PassOptionsToPackage{hyphens}{url}
\PassOptionsToPackage{dvipsnames,svgnames,x11names}{xcolor}
\documentclass[
  11pt,
]{article}
\usepackage{xcolor}
\usepackage{amsmath,amssymb}
  \usepackage[T1]{fontenc}
  \usepackage[utf8]{inputenc}
  \usepackage{textcomp} 
\usepackage{lmodern}
\IfFileExists{upquote.sty}{\usepackage{upquote}}{}
\IfFileExists{microtype.sty}{
  \usepackage[]{microtype}
  \UseMicrotypeSet[protrusion]{basicmath} 
}{}
\makeatletter
\@ifundefined{KOMAClassName}{
  \IfFileExists{parskip.sty}{%
    \usepackage{parskip}
  }{
    \setlength{\parindent}{0pt}
    \setlength{\parskip}{6pt plus 2pt minus 1pt}}
}{
  \KOMAoptions{parskip=half}}
\makeatother
\usepackage{longtable,booktabs,array}
\usepackage{calc} 
\usepackage{etoolbox}
\makeatletter
\patchcmd\longtable{\par}{\if@noskipsec\mbox{}\fi\par}{}{}
\makeatother
\IfFileExists{footnotehyper.sty}{\usepackage{footnotehyper}}{\usepackage{footnote}}
\makesavenoteenv{longtable}
\usepackage{graphicx}
\makeatletter
\newsavebox\pandoc@box
\newcommand*\pandocbounded[1]{
  \sbox\pandoc@box{#1}%
  \Gscale@div\@tempa{\textheight}{\dimexpr\ht\pandoc@box+\dp\pandoc@box\relax}%
  \Gscale@div\@tempb{\linewidth}{\wd\pandoc@box}%
  \ifdim\@tempb\p@<\@tempa\p@\let\@tempa\@tempb\fi
  \ifdim\@tempa\p@<\p@\scalebox{\@tempa}{\usebox\pandoc@box}%
  \else\usebox{\pandoc@box}%
  \fi%
}
\def\fps@figure{htbp}
\makeatother
\providecommand{\tightlist}{%
  \setlength{\itemsep}{0pt}\setlength{\parskip}{0pt}}

\usepackage{newpxtext}               
\usepackage{newpxmath}               
\usepackage[scaled=0.92]{helvet}     
\usepackage{microtype}

\usepackage[letterpaper,margin=1in]{geometry}
\usepackage{setspace}
\usepackage{parskip}                 
\usepackage{etoolbox}
\usepackage{xurl}                    

\usepackage[perpage]{footmisc}

\usepackage{endnotes}

\usepackage{float}
\usepackage{caption}
\AtBeginEnvironment{longtable}{\small}

\usepackage{titlesec}
\titleformat{\section}{\sffamily\large\bfseries}{\thesection.}{0.5em}{}
\titleformat{\subsection}{\sffamily\normalsize\bfseries}{\thesubsection}{0.5em}{}
\titlespacing*{\section}{0pt}{1.3em}{0.5em}
\titlespacing*{\subsection}{0pt}{0.9em}{0.4em}

\AtBeginEnvironment{quote}{\small}
\usepackage{bookmark}
\IfFileExists{xurl.sty}{\usepackage{xurl}}{} 
\hypersetup{
  pdftitle={Quantum Horizon},
  colorlinks=true,
  linkcolor={black},
  filecolor={Maroon},
  citecolor={black},
  urlcolor={black},
  pdfcreator={LaTeX via pandoc}}

\title{Quantum Horizon}
\usepackage{etoolbox}
\makeatletter
\providecommand{\subtitle}[1]{
  \apptocmd{\@title}{\par {\large #1 \par}}{}{}
}
\makeatother
\subtitle{An evaluation of quantum computing as a threat to Bitcoin and
Ethereum}
\author{Iosif M. Gershteyn\thanks{ImmuVia; Ajax Biomedical Foundation.}{}\hspace{0.45em}\& Jacob A. Alber\thanks{Sataresse AI.}}
\date{June 4, 2026}

\begin{document}
\maketitle
\begin{abstract}
Quantum computing poses a real, broad-based, but bounded and
substantially mitigable threat to Bitcoin and Ethereum. We separate the
two quantum algorithms that public discussion routinely conflates:
Shor's algorithm breaks the elliptic-curve signatures (ECDSA over
secp256k1, BLS over BLS12-381) that authorize spending, whereas Grover's
algorithm does not meaningfully threaten proof-of-work mining, which is
protected by a merely quadratic speedup, fault-tolerant per-operation
costs, a square-root parallelization wall, and difficulty adjustment.
Folding hardware scaling, the falling resource requirement, a
fault-tolerance readiness lag, and expert surveys into a single
Monte-Carlo forecast yields a wide, bimodal arrival distribution for a
cryptographically relevant quantum computer: about a one-in-six chance
by 2035, near 30\% by 2040, and about 60\% by 2050. Exposure is
concentrated and mostly migratable: of Bitcoin's roughly six million
quantum-exposed coins only about 2.3 million are irreducibly at risk,
while 50 to 65\% of Ether sits at key-revealed accounts that can adopt
post-quantum signatures. A timely migration beats even an optimistic
2035 machine, so the binding constraint is governance, not technology. A
survey of the top twenty cryptocurrencies finds none fully post-quantum.
Reproducible models accompany every quantitative claim.
\end{abstract}

\subsection{Executive summary}\label{executive-summary}

Quantum computing is a \textbf{real, broad-based, but bounded and
substantially mitigable} threat to Bitcoin and Ethereum, and recent
results are \textbf{compressing} the timeline. The four core findings:

\begin{enumerate}
\def\labelenumi{\arabic{enumi}.}
\tightlist
\item
  \textbf{No cryptographically-relevant quantum computer (CRQC) exists
  today}, and the gap to one is large. Breaking the elliptic-curve
  cryptography that secures both chains requires on the order of
  \textbf{1,200--2,330 logical qubits, or 0.5--320 million physical
  qubits}, against the roughly 1,000--1,200 physical qubits (and at most
  about a hundred logical qubits) on 2026's best machines. That is a gap
  of about 400--500$\times$ even versus the \emph{most aggressive} 2026
  estimate, and orders of magnitude larger versus conservative ones, on
  top of a fault-tolerant hardware stack that does not yet exist.
\item
  \textbf{The threat is to \emph{signatures}, not to mining.} Shor's
  algorithm can break the signature schemes (ECDSA over secp256k1, and
  BLS over BLS12-381). Grover's algorithm does \textbf{not} meaningfully
  threaten Bitcoin's proof-of-work: its speedup is only quadratic, it is
  crushed by fault-tolerant per-operation cost and a $\sqrt{K}$
  parallelization wall, and difficulty adjustment cancels what remains.
\item
  \textbf{Exposure is bounded and mostly migratable.} Of Bitcoin's
  roughly 6 million quantum-exposed coins (about 30\% of supply), only
  \textbf{2.3 million (about 12\%) are irreducibly at risk}: dormant,
  lost, or Satoshi-era coins whose owners can never move them.
  Ethereum's at-rest exposure is broader, around 50--65\% of ETH sits at
  used accounts (a measurement-anchored figure; see §5.1), but it is
  more readily migratable.
\item
  \textbf{Migration is feasible, and the binding constraint is
  governance, not technology.} Post-quantum standards already
  exist,\endnote{National Institute of Standards and Technology, ``NIST Releases First 3
  Finalized Post-Quantum Encryption Standards'' (FIPS 203 ML-KEM, FIPS 204
  ML-DSA, FIPS 205 SLH-DSA), August 13, 2024.
  \url{https://www.nist.gov/news-events/news/2024/08/nist-releases-first-3-finalized-post-quantum-encryption-standards}}
  and both chains have concrete migration paths. A timely start beats
  even an optimistic 2035 CRQC with years to spare; only a
  \emph{severely delayed} migration loses the race.
\end{enumerate}

Two further results frame the rest of the paper. Combining hardware
scaling, the falling resource requirement, a fault-tolerance readiness
lag, and expert surveys into one forecast yields a wide,
\textbf{bimodal} distribution rather than a tidy point estimate: about a
\textbf{one-in-six chance of a CRQC by 2035, near 30\% by 2040, and
about 60\% by 2050}, across an 80\% range of roughly 2032 to 2060
(§3.2). And a supporting survey of the \textbf{top 20 cryptocurrencies}
finds that \textbf{none is fully post-quantum on its main signing path},
with Bitcoin and Ethereum near the \emph{less-ready} end of the major
chains; readiness turns far more on a coin's exposure model and
migration program than on its choice of curve (§6).

\begin{longtable}[]{@{}
  >{\raggedright\arraybackslash}p{(\linewidth - 2\tabcolsep) * \real{0.3472}}
  >{\raggedright\arraybackslash}p{(\linewidth - 2\tabcolsep) * \real{0.6528}}@{}}
\caption{\textbf{Table 1.} Key numbers at a glance; each is developed
and sourced in the body (§2--§7).}\tabularnewline
\toprule\noalign{}
\begin{minipage}[b]{\linewidth}\raggedright
Key number
\end{minipage} & \begin{minipage}[b]{\linewidth}\raggedright
Value
\end{minipage} \\
\midrule\noalign{}
\endfirsthead
\toprule\noalign{}
\begin{minipage}[b]{\linewidth}\raggedright
Key number
\end{minipage} & \begin{minipage}[b]{\linewidth}\raggedright
Value
\end{minipage} \\
\midrule\noalign{}
\endhead
\bottomrule\noalign{}
\endlastfoot
Breaking secp256k1 requires & \textasciitilde1,200--2,330 logical
qubits; \textasciitilde0.5--320 million physical \\
Best demonstrated hardware (2026) & \textasciitilde1,000--1,200 physical
qubits; at most \textasciitilde100 logical \\
Odds of a CRQC & \textasciitilde1-in-6 by 2035 (8--24\% across
weightings); \textasciitilde30\% by 2040; \textasciitilde60\% by 2050 \\
Bitcoin exposed at rest & \textasciitilde6.0M BTC (\textasciitilde30\%
of supply): \textasciitilde2.3M irreducible + \textasciitilde3.7M
migratable \\
Ether at used accounts & 50--65\% of supply, most defensibly 55--60\% \\
Quantum mining & \textasciitilde21 TH/s per machine at an optimistic 100
GHz, about one-tenth of one ASIC \\
The migration race & a prompt 2026 start finishes
\textasciitilde2029--2031, beating even an optimistic 2035 CRQC \\
\end{longtable}

The window is finite and shrinking. Two \emph{distinct} 2025--26 results
moved the goalposts. An algorithmic advance cut the cost of factoring
RSA-2048 roughly
twentyfold,\endnote{Craig Gidney, ``How to factor 2048 bit RSA integers with less than a
million noisy qubits,'' arXiv:2505.15917 (2025).
\url{https://arxiv.org/abs/2505.15917}} and a separate Google and
Ethereum Foundation whitepaper put breaking Bitcoin's secp256k1 curve at
under half a million physical
qubits.\endnote{Ryan Babbush, Adam Zalcman, Craig Gidney, Michael Broughton, Tanuj
Khattar, Hartmut Neven, Thiago Bergamaschi, Justin Drake, Dan Boneh
(Google Quantum AI, Ethereum Foundation, Stanford), ``Securing Elliptic
Curve Cryptocurrencies against Quantum Vulnerabilities: Resource
Estimates and Mitigations,'' arXiv:2603.28846 (2026).
\url{https://arxiv.org/abs/2603.28846}} Bottom-up hardware models and
expert surveys still disagree on the date by roughly a decade, and the
systemic forecast spans both rather than choosing between them. The
threat is uncertain in timing, with a real chance inside the next decade
and a center of mass in the 2040s, and no longer comfortably distant,
which makes the case for acting now stronger, not weaker. The right
posture is prepared urgency, not alarm.

\begin{quote}
\textbf{General Summary.} Quantum computers could one day forge the
digital signatures that let people spend Bitcoin and Ethereum. No such
machine exists yet, and building one is still years away, but the
estimates of how hard it will be have dropped sharply in the past year.
The danger is real but limited and fixable, and the smart move is to
start upgrading now rather than panic.
\end{quote}

\begin{center}\rule{0.5\linewidth}{0.5pt}\end{center}

\subsection{1. Scope and method}\label{scope-and-method}

Every load-bearing number is tied to a primary or cross-checked source,
cited in the numbered endnotes collected at the end of the paper, and
every quantitative claim is backed by a small, reproducible calculation
or model. (Lettered footnotes at the bottom of each page serve a
different purpose: brief definitions of technical terms at first
mention.) Numbers are given as \textbf{ranges with stated assumptions},
not as point predictions. The models were built so that independent ones
could check each other, and where two methods disagree we present the
disagreement rather than average it away.

The paper is positioned against, and indebted to, several prior
analyses: the Google, Ethereum Foundation, and Stanford resource
estimates and mitigations,\endnotemark[3] Deloitte's
Bitcoin\endnote{Itan Barmes, Bram Bosch, Olaf Haalstra, ``Quantum computers and the
Bitcoin blockchain,'' Deloitte Netherlands (2020).
\url{https://www.deloitte.com/nl/en/services/consulting-risk/perspectives/quantum-computers-and-the-bitcoin-blockchain.html}}
and
Ethereum\endnote{Itan Barmes, Bram Bosch, Olaf Haalstra, ``Quantum risk to the Ethereum
blockchain,'' Deloitte Netherlands (2021).
\url{https://www.deloitte.com/nl/en/services/consulting-risk/perspectives/quantum-risk-to-the-ethereum-blockchain.html}}
exposure scans, CoinShares' Bitcoin
assessment,\endnote{Chris Bendiksen, ``Quantum Vulnerability in Bitcoin: A Manageable
Risk,'' CoinShares Research (February 2026).
\url{https://coinshares.com/insights/research-data/quantum-vulnerability-in-bitcoin-a-manageable-risk/}}
and the Global Risk Institute's expert
surveys.\endnote{Michele Mosca and Marco Piani, ``Quantum Threat Timeline Report,''
Global Risk Institute and evolutionQ (2024 and 2025 editions).
\url{https://globalriskinstitute.org/publication/2024-quantum-threat-timeline-report/}
and
\url{https://globalriskinstitute.org/publication/quantum-threat-timeline-report-2025b/}}
What it adds is integration and arbitration: a systemic break-year
forecast that folds hardware scaling, the falling resource requirement,
fault-tolerance readiness, and the surveys into one distribution, and
reports that distribution's bimodality rather than averaging it away
(§3.2); the separation of exposed supply into irreducible versus
migratable (§4.2); a calibrated mining model that retires the
proof-of-work confusion (§2.2); a migration-race reading of Mosca's
inequality (§7.4); and the cross-market readiness survey (§6).

The deep analysis covers the base layers of Bitcoin and Ethereum,
including Ethereum's consensus-layer BLS signatures and its KZG
data-availability commitments. Section 6 widens the lens to the top 20
cryptocurrencies by market value, to place Bitcoin and Ethereum in
context. Layer-2 networks appear only where they change the base-layer
picture. All market, chain, and hardware figures are as of early June
2026 unless dated otherwise.

\begin{quote}
\textbf{General Summary.} Everything below is sourced and backed by a
calculation a reader could redo. Where the evidence cuts against the
paper's thesis, the paper says so.
\end{quote}

\begin{center}\rule{0.5\linewidth}{0.5pt}\end{center}

\subsection{2. What quantum actually breaks, and what it does
not}\label{what-quantum-actually-breaks-and-what-it-does-not}

Two quantum algorithms matter, and conflating them is the single most
common error in this debate.

\subsubsection{2.1 Shor's algorithm versus digital signatures (the real
threat)}\label{shors-algorithm-versus-digital-signatures-the-real-threat}

Bitcoin and Ethereum authorize spending with \textbf{ECDSA over the
secp256k1
curve}\footnote{ECDSA (the Elliptic Curve Digital Signature Algorithm), here on the curve named secp256k1, is the method Bitcoin and Ethereum use to prove that a transaction was authorized by the owner of the coins.},
and Ethereum's consensus layer uses \textbf{BLS signatures over
BLS12-381}\footnote{BLS is a second signature scheme, used on the curve BLS12-381 by Ethereum's validators (the nodes that run its proof-of-stake consensus). Like ECDSA, it can be broken by a quantum computer.}.
Both rest on the elliptic-curve discrete-logarithm
problem.\footnote{The hard math problem these signatures rely on. Each user has a private key (kept secret) and a matching public key (shared openly); recovering the private key from the public key would require solving this problem, which ordinary computers cannot but a large quantum computer could.}
\textbf{Shor's algorithm solves that problem
efficiently}\footnote{Shor's algorithm (Peter Shor, 1994) is the quantum method that solves that problem efficiently, letting a sufficiently large, error-corrected quantum computer derive a private key from a public one. Such a machine is the cryptographically-relevant quantum computer, or CRQC, referred to throughout.}
on a large enough quantum computer, so a CRQC can recover a private key
from a \emph{public key}. The whole question of ``how exposed is a
coin?'' therefore reduces to ``is its public key visible on the
blockchain?'' (see §4 and §5).

One subtlety, made precise by the 2026 Google and Ethereum Foundation
whitepaper, runs through this paper: \textbf{BLS12-381 is somewhat
\emph{harder} to break than secp256k1, not easier.} Shor's cost scales
with the size of the field in which the curve arithmetic happens, and
BLS12-381 uses 381-bit coordinates against secp256k1's 256-bit. That is
``a 50\% increase in the size of some quantum registers'' and implies
``a somewhat larger CRQC,'' though the premium is ``modest,'' and the
consensus layer ``should be considered at-rest vulnerable to the same
first-generation CRQCs.''\endnotemark[3]

\subsubsection{2.2 Grover's algorithm versus hashes and proof-of-work
(not the
threat)}\label{grovers-algorithm-versus-hashes-and-proof-of-work-not-the-threat}

Grover's
algorithm\footnote{Grover's algorithm is the best known quantum attack on brute-force search. It offers only a square-root speedup, far too weak to threaten mining or the hash functions below.}
gives only a \textbf{quadratic} speedup on unstructured search. Against
SHA-256, Keccak-256, and the Bitcoin address hash
HASH160\footnote{SHA-256, Keccak-256 and HASH160 are hash functions: one-way fingerprinting functions. SHA-256 underpins Bitcoin mining, while HASH160 and Keccak-256 form Bitcoin and Ethereum addresses. Quantum computers weaken them only slightly.},
it cuts preimage security from 256 to about 128 bits (and 160 to about
80 for HASH160, the thinnest margin of the three, though still
Grover-bounded and reachable only if an address is reused). The 128-bit
cases remain intractable, and in practice the attack is far worse for
the attacker than that suggests: a
fault-tolerant\footnote{Real quantum hardware is noisy. A fault-tolerant machine uses heavy error-correction to compute reliably regardless; building one at the scale these attacks need is the central unsolved engineering problem.}
SHA-256 preimage attack costs ``approximately 2$^{166}$
logical-qubit-cycles'' and ``may be as much as 275 billion times more
expensive than one would expect from the simple query
analysis.''\endnote{Matthew Amy, Olivia Di Matteo, Vlad Gheorghiu, Michele Mosca, Alex
Parent, John Schanck, ``Estimating the cost of generic quantum pre-image
attacks on SHA-2 and SHA-3,'' SAC 2016, arXiv:1603.09383.
\url{https://arxiv.org/abs/1603.09383}} Hashing is not the weak link.

The most consequential version of this confusion is the claim that
``quantum computers will out-mine
Bitcoin.''\footnote{Bitcoin mining, called proof-of-work, secures the network: machines race to find an input whose hash falls below a target. That target's difficulty adjusts automatically to keep blocks arriving about every ten minutes, which neutralizes any single faster miner.}
They will not. Our mining-competitiveness model, calibrated to the 2017
Aggarwal et al.~benchmark of 13.8 GH/s at a 66.7 MHz gate
speed,\endnote{Divesh Aggarwal, Gavin Brennen, Troy Lee, Miklos Santha, Marco
Tomamichel, ``Quantum attacks on Bitcoin, and how to protect against
them,'' \emph{Ledger} (2018), arXiv:1710.10377.
\url{https://arxiv.org/abs/1710.10377}} shows why.

\begin{itemize}
\tightlist
\item
  Bitcoin's June-2026 difficulty implies about 2$^{79}$ hashes per
  block. Grover would need roughly 2$^{40}$ iterations, but each
  iteration is a fault-tolerant, reversible double-SHA-256 oracle of
  474,168 T-gates dominated by magic-state distillation, not a cheap
  ASIC hash.\endnotemark[9]
\item
  A single quantum machine, even at an optimistic 100 GHz gate speed,
  reaches only about 21 TH/s, roughly one-tenth of a single modern ASIC,
  against a network running near 860 EH/s (about 4.3 million
  ASIC-equivalents; the count tracks the $\pm$20\% hashrate-estimate
  spread).
\item
  Grover barely parallelizes. K machines buy only a $\sqrt{K}$
  speedup,\endnote{Christof Zalka, ``Grover's quantum searching algorithm is optimal,''
  \emph{Physical Review A} 60 (1999), arXiv:quant-ph/9711070.
  \url{https://arxiv.org/abs/quant-ph/9711070}} against linear scaling
  for ASIC farms. Reaching 51\% of the network would take on the order
  of 7$\times$10$^{13}$ quantum machines at 100 GHz.
\item
  Even if a quantum miner did gain an edge, \textbf{difficulty
  adjustment absorbs it}. The miner simply becomes one more participant,
  not a break of proof-of-work.
\end{itemize}

\begin{figure}
\centering
\includegraphics[width=5.5in,height=\textheight,keepaspectratio,alt={Figure 1. Effective quantum ``hashrate'' versus gate speed, compared with one ASIC and the whole 2026 network.}]{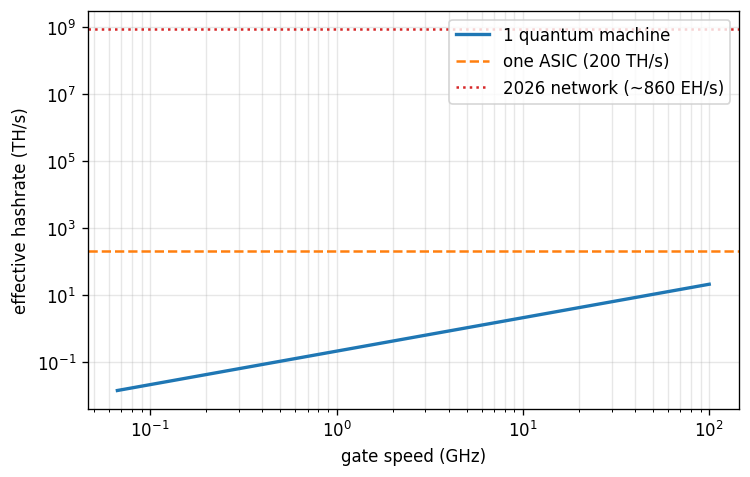}
\caption{\textbf{Figure 1.} Effective quantum ``hashrate'' versus gate
speed, compared with one ASIC and the whole 2026 network.}
\end{figure}

\textbf{Verdict: proof-of-work is not the weak point.} The only residual
mining concern is ordinary economics, the centralization or 51\%-style
risk that would arise if quantum mining ever became cheap and
concentrated, which the parallelization wall makes implausible for the
foreseeable future.

\begin{quote}
\textbf{General Summary.} There are two different quantum attacks, and
people constantly mix them up. One (Shor's) can forge the signatures
that authorize spending, and that is the real danger. The other
(Grover's) only speeds up the guessing game behind mining, and nowhere
near enough to matter: a quantum miner would be thousands of times
slower than today's specialized hardware, and the network would simply
adjust. Quantum threatens the locks, not the mining.
\end{quote}

\begin{center}\rule{0.5\linewidth}{0.5pt}\end{center}

\subsection{3. The timeline: compressing, but no machine
today}\label{the-timeline-compressing-but-no-machine-today}

\subsubsection{3.1 What it takes, and what
exists}\label{what-it-takes-and-what-exists}

Estimates of the difficulty of breaking secp256k1 have fallen sharply as
algorithms improved, with no change in hardware. They have come down
from Roetteler et al.~in 2017 (about 2,330 logical
qubits\footnote{A qubit is a quantum bit. Real (physical) qubits are noisy, so many are bundled together with error-correction to form one reliable logical qubit. Breaking these signatures needs on the order of a thousand logical qubits, which in turn requires hundreds of thousands to millions of physical ones.}
and 2.3$\times$10$^{11}$ Toffoli
gates)\endnote{Martin Roetteler, Michael Naehrig, Krysta Svore, Kristin Lauter,
``Quantum resource estimates for computing elliptic curve discrete
logarithms,'' ASIACRYPT 2017, arXiv:1706.06752.
\url{https://arxiv.org/abs/1706.06752}} to a 2026 frontier of
\textbf{1,200--1,450 logical qubits, 70--90 million Toffoli gates, and
fewer than 500,000 physical qubits} (surface code, 10$^{-3}$ error rate)
in the Google, Ethereum Foundation, and Stanford
whitepaper.\endnotemark[3] Conservative estimates remain far higher;
Webber et al.~in 2022 required about 317 million physical qubits to
break the curve within an
hour.\endnote{Mark Webber, Vincent Elfving, Sebastian Weidt, Winfried Hensinger, ``The
impact of hardware specifications on reaching quantum advantage in the
fault tolerant regime,'' \emph{AVS Quantum Science} 4 (2022),
arXiv:2108.12371. \url{https://arxiv.org/abs/2108.12371}}

Against those requirements, 2026 hardware offers roughly 1,000--1,200
\emph{physical} qubits (IBM's 1,121-qubit Condor; Atom Computing's
1,180-qubit
array)\endnote{IBM's 1,121-qubit Condor processor (December 2023), see
\url{https://en.wikipedia.org/wiki/IBM_Condor}; Atom Computing's
1,225-site neutral-atom array, populated with 1,180 qubits (October
2023).
\url{https://www.prnewswire.com/news-releases/quantum-startup-atom-computing-first-to-exceed-1-000-qubits-301964712.html}}
and demonstrated logical-qubit counts only in the tens to roughly a
hundred (Google's below-threshold 105-qubit Willow chip, the 48 logical
qubits on Quantinuum's Helios, and up to 96 logical qubits in a Harvard,
MIT, and QuEra
system).\endnote{Google, ``Meet Willow, our state-of-the-art quantum chip'' (December
2024),
\url{https://blog.google/technology/research/google-willow-quantum-chip/};
Quantinuum, Helios commercial launch with 48 logical qubits on 98
physical (November 2025),
\url{https://www.quantinuum.com/press-releases/quantinuum-announces-commercial-launch-of-new-helios-quantum-computer-that-offers-unprecedented-accuracy-to-enable-generative-quantum-ai-genqai};
Harvard, MIT, and QuEra collaborators, fault-tolerant architecture with
up to 96 logical qubits on 448 neutral atoms, \emph{Nature} (November
2025), see
\url{https://quantumcomputingreport.com/harvard-and-collaborators-demonstrate-scalable-fault-tolerant-architecture-with-448-neutral-atom-qubits/}}
\textbf{No CRQC exists, and every aggressive 2026 source says so
explicitly}; the whitepaper itself states that ``cryptographically
relevant quantum computers do not exist as of today.''\endnotemark[3]

\subsubsection{3.2 A systemic forecast for
``when''}\label{a-systemic-forecast-for-when}

Rather than rely on a single line of evidence, we fold four signals into
one
Monte-Carlo\footnote{Monte-Carlo: a method that estimates a range of outcomes by running a model many thousands of times with randomly varied inputs and tallying the results.}
break-year forecast:

\begin{itemize}
\tightlist
\item
  the \textbf{resource requirement} and its \textbf{algorithmic decline}
  over time (the roughly 20$\times$ drop in RSA estimates, and the
  parallel fall in elliptic-curve estimates, mean the target is moving,
  so we let the requirement halve every 4 to 20 years);
\item
  \textbf{hardware scaling} (best-device physical-qubit count doubling
  every 1.0 to 2.5 years);
\item
  a \textbf{fault-tolerance readiness lag} of 2 to 12 years, the gap
  between having enough raw physical qubits and operating a working
  fault-tolerant machine at cryptanalytic scale (a step the naive
  bottom-up estimate omits, which is why it reads as an optimistic lower
  bound); and
\item
  the \textbf{expert surveys} (Global Risk Institute and Mosca) as an
  independent estimator,\endnotemark[7] blended in at equal weight.
\end{itemize}

The two estimators disagree by more than a decade: the expert surveys
center near 2038--2040, while bottom-up physics (with the
requirement-decline and the readiness lag) centers near 2052. Because
they disagree, \textbf{the combined distribution is
bimodal}\footnote{A bimodal distribution has two separate peaks instead of one. Here an earlier survey-based peak and a later physics-based peak leave the average stranded in an unlikely valley between them.},
with a near-term survey hump and a later physics hump (Figure 2), and
its single median falls in the low-probability trough between them near
2046. We therefore read the forecast off its cumulative probabilities
and its spread, not off the median.

\begin{figure}
\centering
\includegraphics[width=5.5in,height=\textheight,keepaspectratio,alt={Figure 2. The two estimators (survey and physics) and the bimodal combined distribution; the median falls in the trough between the modes.}]{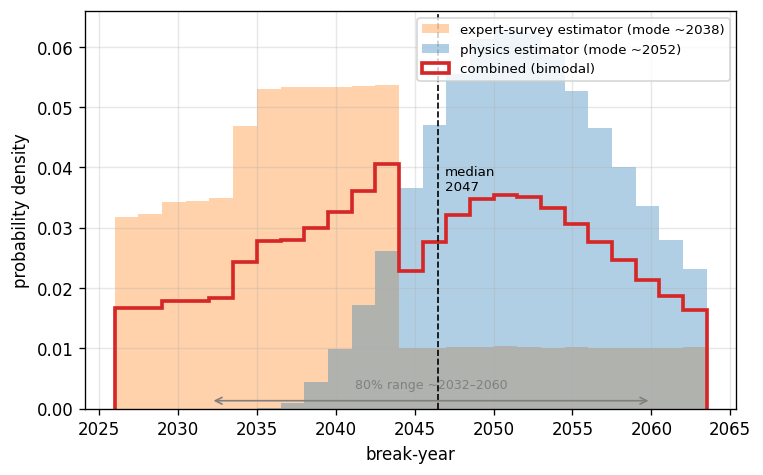}
\caption{\textbf{Figure 2.} The two estimators (survey and physics) and
the bimodal combined distribution; the median falls in the trough
between the modes.}
\end{figure}

The combined forecast assigns about a \textbf{one-in-six chance of a
CRQC by 2035, near 30\% by 2040, and about 60\% by 2050}, across an
\textbf{80\% range of roughly 2032 to 2060.} Two caveats keep this from
being read as more than it is. It blends two different \emph{kinds} of
evidence, a physics model and a subjective expert poll, rather than
cross-validating one against the other. It weights them equally for want
of a principled reason to privilege either: the bottom-up model the
paper itself calls an optimistic lower bound, or the surveyed experts.
That weight does real work: at survey weights from 0.25 to 0.75 the
by-2035 probability runs from about 8\% to 24\% (and the median from
about 2050 to 2041), so the near-term tail tracks the weighting rather
than being a hard result. The largest sources of \emph{spread},
separately, are the hardware doubling time and the resource requirement;
the fault-tolerance lag mainly moves the later mode (removing it leaves
the by-2035 probability essentially unchanged), and the
algorithmic-decline rate shifts the center while adding little width.

The forecast is a forecast, not a prediction, and the bands are wide on
purpose. The decision-relevant point survives the caveats: across the
full plausible weighting, the chance of a CRQC by 2035 is somewhere
between roughly one-in-twelve and one-in-four, a near-term tail large
enough to start migrating now (§7.4 shows a prompt start still wins the
race) but far from a basis for alarm. Neither ``comfortably distant''
nor ``panic now'' survives contact with the evidence.

\begin{quote}
\textbf{General Summary.} A machine big enough to break these signatures
does not exist, and the experts and the hardware models disagree by more
than a decade about when one will arrive: roughly the late 2030s versus
the early 2050s. Taken together the evidence gives a wide spread, with
about a one-in-six chance by 2035 and better-than-even odds by 2050.
Soon enough to prepare for, not soon enough to panic about.
\end{quote}

\begin{center}\rule{0.5\linewidth}{0.5pt}\end{center}

\subsection{4. Bitcoin's exposure}\label{bitcoins-exposure}

\subsubsection{4.1 Where the public key is, by script
type}\label{where-the-public-key-is-by-script-type}

A Bitcoin coin is quantum-exposed exactly when its public key is visible
on the blockchain.\endnotemark[9]

\begin{longtable}[]{@{}
  >{\raggedright\arraybackslash}p{(\linewidth - 4\tabcolsep) * \real{0.2192}}
  >{\raggedright\arraybackslash}p{(\linewidth - 4\tabcolsep) * \real{0.3699}}
  >{\raggedright\arraybackslash}p{(\linewidth - 4\tabcolsep) * \real{0.4110}}@{}}
\caption{\textbf{Table 2.} Bitcoin script types and whether they expose
the public key at rest.}\tabularnewline
\toprule\noalign{}
\begin{minipage}[b]{\linewidth}\raggedright
Script type
\end{minipage} & \begin{minipage}[b]{\linewidth}\raggedright
What is on-chain
\end{minipage} & \begin{minipage}[b]{\linewidth}\raggedright
Public key exposed at rest?
\end{minipage} \\
\midrule\noalign{}
\endfirsthead
\toprule\noalign{}
\begin{minipage}[b]{\linewidth}\raggedright
Script type
\end{minipage} & \begin{minipage}[b]{\linewidth}\raggedright
What is on-chain
\end{minipage} & \begin{minipage}[b]{\linewidth}\raggedright
Public key exposed at rest?
\end{minipage} \\
\midrule\noalign{}
\endhead
\bottomrule\noalign{}
\endlastfoot
\textbf{P2PK}, \textbf{P2MS} & the public key(s) themselves &
\textbf{Yes, always} \\
\textbf{P2PKH / P2WPKH} & a hash (HASH160) of the public key & No, until
first spend; then \textbf{permanent if reused} \\
\textbf{P2SH / P2WSH} & a hash of a script & Only when the script
reveals a key \\
\textbf{P2TR (Taproot)} & a raw 32-byte public key & \textbf{Yes
(key-path)}, a regression despite being the newest type \\
\end{longtable}

The behavioral driver is address
\textbf{reuse}.\footnote{Reusing an address means spending from it more than once. The first spend publishes its public key permanently, so any coins later sent back to that same address are exposed.}
Once an address has spent, its key is public forever, so any remaining
or future balance there is exposed at rest.

\subsubsection{4.2 How much is at risk, and the distinction that
matters}\label{how-much-is-at-risk-and-the-distinction-that-matters}

Our Bitcoin-exposure analysis reconciles several independent chain
analyses: the three 2025--26 measurements (Glassnode at 6.04M, or 30.2\%
of
supply;\endnote{``Measuring Bitcoin's Quantum-Exposed Supply,'' Glassnode Research (May
2026).
\url{https://research.glassnode.com/measuring-bitcoins-quantum-exposed-supply/}}
Coinbase at about 6.9M; CoinDesk at about 7M) agree to within about five
percentage points of supply, and Deloitte's older 2020 scan, at about
25\%, sits just below them.\endnotemark[4] But ``6 million exposed''
conflates two different things.

\begin{itemize}
\tightlist
\item
  \textbf{Irreducible (about 2.3 million BTC, 12\%):} coins that are
  \emph{dormant} across all script types, including but not limited to
  the roughly 1 million Satoshi-era coins, whose owners can never move
  them to safety.\endnotemark[3] This is a \emph{different axis} from
  the roughly 1.7 million held in the P2PK format: being in P2PK means
  \emph{exposed}, not necessarily \emph{lost}, and most non-Satoshi P2PK
  keys may still be controlled and are therefore migratable. The
  irreducible figure is the \emph{dormant} subset specifically, and it
  is the hard floor of permanent risk.
\item
  \textbf{Migratable (about 3.7 million BTC, 19\%):} exposed but
  spendable. Owners \emph{can} sweep these into quantum-safe outputs
  given warning.
\item
  \textbf{Protected (about 65--70\%):} held in fresh, never-reused,
  hash-based addresses, where the public key is revealed only briefly,
  at the moment of spending.
\end{itemize}

\begin{figure}
\centering
\includegraphics[width=5.5in,height=\textheight,keepaspectratio,alt={Figure 3. Bitcoin's supply split into permanently-at-risk, migratable, and protected coins.}]{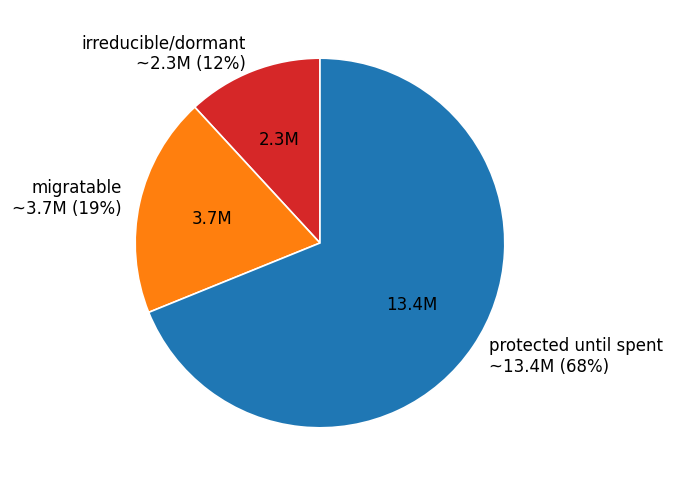}
\caption{\textbf{Figure 3.} Bitcoin's supply split into
permanently-at-risk, migratable, and protected coins.}
\end{figure}

\subsubsection{4.3 The live attack: mempool
sniping}\label{the-live-attack-mempool-sniping}

The 65--70\% of protected coins have one window of vulnerability. When
such a coin is spent, its public key enters the public waiting area (the
mempool). A \textbf{fast-clock} CRQC (a superconducting or photonic
machine, in the whitepaper's key distinction)\endnotemark[3] could, from
a primed state, derive the key in roughly 9--12 minutes and broadcast a
higher-fee replacement transaction before the original confirms. Our
mempool-race model reproduces the literature's 41\% figure as a
\emph{best-case corner} (a 9-minute derivation, zero network
propagation, a single confirmation) and puts the realistic per-spend
success nearer \textbf{30\%, and only if a fast-clock CRQC exists}. The
attack is bounded to the one-block window, and impossible for a
slow-clock machine (trapped-ion or neutral-atom), which cannot derive a
key inside ten minutes. The mitigations are mundane and effective: never
reuse addresses, and migrate.

\begin{quote}
\textbf{General Summary.} Not all Bitcoin is equally at risk. A coin is
only vulnerable once its public key has appeared on the blockchain.
Coins kept in fresh addresses stay hidden behind a one-way fingerprint
and are safe until spent. The unfixable risk is a small slice, about 2.3
million coins or one in eight, that are lost or dormant (including
Satoshi's), which no one can ever move to safety. Mining, and the
everyday coin sitting in a fresh wallet, are not the problem.
\end{quote}

\begin{center}\rule{0.5\linewidth}{0.5pt}\end{center}

\subsection{5. Ethereum's exposure}\label{ethereums-exposure}

\subsubsection{5.1 Broader at rest, by
design}\label{broader-at-rest-by-design}

An Ethereum address is the last 20 bytes of a hash of the public key, so
an unused account's key is hidden. But \textbf{the public key can be
recovered from any ECDSA signature} (via the \texttt{ecrecover}
operation), so \textbf{every account that has ever sent a transaction
has an effectively public key}, and Ethereum's account
model\footnote{Unlike Bitcoin's coin-by-coin design, Ethereum tracks balances in accounts, much like bank balances. An account reveals its public key the first time it sends a transaction, and accounts are reused constantly.}
encourages address reuse. The result is structural over-exposure
relative to Bitcoin's model.

Our Ethereum-exposure analysis places at-rest account exposure at
\textbf{50--65\% of all ETH (about 61--79 million), most defensibly
55--60\%}, and this figure is measurement-anchored rather than analyst
judgment. The often-quoted ``over 65\%'' is Deloitte's 2021 number,
which we confirmed is a real on-chain measurement: a full scan of the
ledger summing the balances of every address that has sent at least one
transaction.\endnotemark[5] Two \emph{independent} 2026 reconstructions
then converge on 55--60\%. A top-down composition (total supply, minus
the roughly 32\% now staked, minus other contract holdings, leaves about
55--63\% in accounts, nearly all of them already revealed) agrees with a
bottom-up on-chain component build (about 45--55\%, once a
Beacon-deposit-contract over-count that naively gave 21\% is
corrected).\endnote{ETH supply and staking figures from ultrasound.money (live dashboard,
2026). \url{https://ultrasound.money/}} The figure has drifted
\emph{down} from 2021's 65\% as roughly a third of supply moved into
staking (a \emph{separate} BLS exposure class; see §5.2) and more into
contracts (which hold no signing key). No clean 2026 re-measurement
exists: the whitepaper declined to compute the Ethereum aggregate as
``very hard to calculate,'' giving only the top 1,000 accounts (20.5
million ETH, about 17\%, a floor),\endnotemark[3] and an independent
June 2026 search for this paper found no fresh public scan either; the
2025--26 reports that quote an Ethereum percentage, including Project
Eleven's May 2026 survey, trace back to the same Deloitte
measurement.\endnote{Project Eleven, ``The Quantum Threat to Blockchains --- 2026 Report''
(May 2026); its Ethereum exposure percentage is sourced to the Deloitte
scan. \url{https://report.projecteleven.com/}} One adjacent quantity
\emph{has} been measured directly: the Ethereum Foundation's
post-quantum team puts the \emph{dormant} exposed slice, supply that is
both quantum-exposed and abandoned, at only about 0.1\%, Ethereum's
counterpart to Bitcoin's 2.3 million irreducible
coins.\endnote{``Post-quantum cryptography on Ethereum,'' ethereum.org, and the
Ethereum Foundation post-quantum team site (2026).
\url{https://ethereum.org/roadmap/future-proofing/quantum-resistance/}
and \url{https://pq.ethereum.org/}} A direct recount of the at-rest
aggregate on current ledger data would collapse the 50--65\% band to a
single number, and it remains the one outstanding data task.

\subsubsection{5.2 Consensus and data
availability}\label{consensus-and-data-availability}

\begin{itemize}
\tightlist
\item
  \textbf{Validators} sign with BLS over BLS12-381, which Shor breaks
  (and which is somewhat harder than secp256k1, per §2.1). About
  37--39.6 million ETH is staked across roughly 900,000 validators (2026
  estimates vary, and after the MaxEB upgrade that count is of validator
  \emph{units}, not distinct operators). Forging a supermajority is
  \textbf{slow}: even at secp256k1 speed, ``20 such machines would need
  more than nine months'' to derive a two-thirds
  supermajority.\endnotemark[3] This is an integrity and finality risk,
  not instant theft, and staked principal is further buffered by
  withdrawal credentials and slashing.
\item
  \textbf{KZG
  commitments}\footnote{A KZG commitment is a small cryptographic proof that a chunk of data was published; Ethereum uses it to support rollups. It is quantum-breakable, but the risk is to short-lived data availability, not to anyone's funds.}
  (the EIP-4844 ``blobs'') are pairing-based and quantum-breakable, but
  the exposure is one of data-availability and soundness, not theft, and
  it is short-lived, since blobs are pruned after about 18 days.
\end{itemize}

\subsubsection{5.3 The asymmetry}\label{the-asymmetry}

Ethereum's exposure is \textbf{broader} than Bitcoin's but more
\textbf{tractable} to fix. Account
abstraction\footnote{Account abstraction lets an Ethereum account be governed by custom rules rather than one fixed signature type, so each account can switch to a quantum-safe scheme on its own schedule.}
(the EIP-7702 and EIP-8141
mechanisms)\endnote{EIP-8141, ``Frame Transaction'' (drafted January 2026; under
consideration for the Hegotá fork), Ethereum Improvement Proposals; see
also EIP-7702. \url{https://eips.ethereum.org/EIPS/eip-8141}} lets each
account adopt a post-quantum signature scheme on its own schedule, and
there is an official Ethereum Foundation post-quantum
roadmap.\endnotemark[18] Bitcoin is the inverse: narrower exposure, but
harder migration (§7). That asymmetry, rather than a single verdict on
which chain is ``more at risk,'' is the accurate picture.

\begin{quote}
\textbf{General Summary.} Ethereum exposes more of its coins by default,
because every account that has ever sent a transaction reveals its key,
and people reuse accounts. Somewhere between half and two-thirds of all
ETH sits in such accounts. The upside: Ethereum can let each account
upgrade to quantum-proof signatures individually, so this broad exposure
is also the easiest kind to fix. The ETH locked up by validators is a
separate, slower, and better-protected problem.
\end{quote}

\begin{center}\rule{0.5\linewidth}{0.5pt}\end{center}

\subsection{6. The broader market: a quantum-readiness
ranking}\label{the-broader-market-a-quantum-readiness-ranking}

Bitcoin and Ethereum are not unusual. A survey of the \textbf{top 20
cryptocurrencies by market value}
(mid-2026)\endnote{Top cryptocurrencies by market capitalization, CoinGecko (June 2026
snapshot). \url{https://www.coingecko.com/}} finds that \textbf{every
one of them signs transactions with an elliptic-curve or pairing-based
scheme, and is therefore Shor-vulnerable; none is fully post-quantum on
its main signing path.} What separates them is not the curve but two
other things: how much of the key is exposed before a coin is spent, and
how far along each project's migration program is.

We rate each chain from 1 (no post-quantum effort) to 5 (a post-quantum
scheme is the live default). No top-20 coin reaches 5. The leaders sit
at 4: a credible, dated migration with code already on a test or opt-in
network. A high rating means a chain is \emph{furthest along a credible
plan}, not that it is safe today; none of the top 20 is quantum-safe
now.

\begin{longtable}[]{@{}
  >{\raggedright\arraybackslash}p{(\linewidth - 8\tabcolsep) * \real{0.1429}}
  >{\raggedright\arraybackslash}p{(\linewidth - 8\tabcolsep) * \real{0.2381}}
  >{\raggedright\arraybackslash}p{(\linewidth - 8\tabcolsep) * \real{0.1548}}
  >{\raggedright\arraybackslash}p{(\linewidth - 8\tabcolsep) * \real{0.3571}}
  >{\centering\arraybackslash}p{(\linewidth - 8\tabcolsep) * \real{0.1071}}@{}}
\caption{\textbf{Table 3.} Quantum-readiness of the top-20
cryptocurrencies (mid-2026 snapshot), rated 1 (no post-quantum effort)
to 5 (post-quantum live by default).}\tabularnewline
\toprule\noalign{}
\begin{minipage}[b]{\linewidth}\raggedright
Coin
\end{minipage} & \begin{minipage}[b]{\linewidth}\raggedright
Signature scheme
\end{minipage} & \begin{minipage}[b]{\linewidth}\raggedright
Exposure model
\end{minipage} & \begin{minipage}[b]{\linewidth}\raggedright
Post-quantum status
\end{minipage} & \begin{minipage}[b]{\linewidth}\centering
Rating
\end{minipage} \\
\midrule\noalign{}
\endfirsthead
\toprule\noalign{}
\begin{minipage}[b]{\linewidth}\raggedright
Coin
\end{minipage} & \begin{minipage}[b]{\linewidth}\raggedright
Signature scheme
\end{minipage} & \begin{minipage}[b]{\linewidth}\raggedright
Exposure model
\end{minipage} & \begin{minipage}[b]{\linewidth}\raggedright
Post-quantum status
\end{minipage} & \begin{minipage}[b]{\linewidth}\centering
Rating
\end{minipage} \\
\midrule\noalign{}
\endhead
\bottomrule\noalign{}
\endlastfoot
XRP & Ed25519 /
secp256k1\endnote{``Cryptographic keys,'' XRP Ledger documentation; and XRPL Standards
discussion \#295 on quantum-resistant signatures (2025--2026).
\url{https://xrpl.org/docs/concepts/accounts/cryptographic-keys} and
\url{https://github.com/XRPLF/XRPL-Standards/discussions/295}} &
account, reuse & ML-DSA on devnet, roadmap to full resistance by 2028 &
\textbf{4} \\
Solana &
Ed25519\endnote{``Solana's Quantum Readiness,'' Solana Foundation (2026).
\url{https://solana.com/news/quantum-readiness}} & account, reuse &
opt-in hash-based ``Winternitz Vault'' live; Falcon selected (not yet
shipped) by both clients & \textbf{4} \\
Zcash & zk-SNARK (shielded), ECDSA
(transparent)\endnote{``Zcash to roll out quantum-recoverable wallets within a month, go
quantum-proof by 2027,'' CoinDesk (May 8, 2026).
\url{https://www.coindesk.com/tech/2026/05/08/zcash-to-roll-out-quantum-recoverable-wallets-within-a-month-go-quantum-proof-by-2027}}
& shielded pool hides keys & dated roadmap: quantum-recoverable wallets
2026, fully PQ 2027 & \textbf{4} \\
BNB & secp256k1; BLS12-381
(consensus)\endnote{``BNB Chain Publishes Post-Quantum Cryptography Migration Report for
BSC,'' BNB Chain / Chainwire (May 14, 2026).
\url{https://chainwire.org/2026/05/14/bnb-chain-publishes-research-report-exploring-post-quantum-cryptography-migration-path-for-bsc/}}
& account, reuse & tested ML-DSA + post-quantum consensus aggregation
(report, May 2026) & \textbf{3.5} \\
TRON &
secp256k1\endnote{``Justin Sun: TRON launches post-quantum upgrade initiative,''
Cryptopolitan (April 15, 2026).
\url{https://www.cryptopolitan.com/tron-launches-post-quantum-initiative/}}
& account, reuse & post-quantum initiative; testnet targeted Q2 2026 &
\textbf{3.5} \\
Ethereum & secp256k1; BLS12-381\endnotemark[18] & account, reuse &
most-developed roadmap; account-abstraction migration, core PQ
\textasciitilde2029 & \textbf{3} \\
USDT / USDC / USDS / LINK & host chain's (mostly secp256k1; Ed25519 on
Solana) & inherits host & rely on host migration; centralized issuers
can also freeze and re-mint & \textbf{3} \\
Monero & Ed25519 ring signatures & stealth addresses (one-time keys) &
research working group formed late 2025 & \textbf{2.5} \\
Cardano &
Ed25519\endnote{``Research program to work on hardening Cardano against quantum
computers,'' Input Output Global (February 2018).
\url{https://www.iog.io/news/research-program-to-work-on-hardening-cardano-against-quantum-computers}}
& UTXO, hashed addresses & post-quantum research program; no live scheme
& \textbf{2.5} \\
Bitcoin & secp256k1 + Schnorr & UTXO, hashed addresses & BIP-360 merged
but \textbf{not activated}; \textasciitilde1.6--1.7M raw-key (P2PK-era)
BTC exposed, of \textasciitilde6M total (§4.2)\endnotemark[6] &
\textbf{2} \\
Stellar / Hyperliquid & Ed25519 / secp256k1 & account, reuse & no public
roadmap located & \textbf{2} \\
Dogecoin & secp256k1 & UTXO, heavy reuse & no post-quantum effort &
\textbf{1.5} \\
\end{longtable}

The table condenses the 20 surveyed assets: the stablecoins and tokens
that inherit a host chain are grouped, and three small,
issuer-controlled entries (Canton, Figure HELOC, and the exchange token
LEO) are omitted because they add no cryptography of their own. Figure 5
rates all nineteen of them individually; the twentieth, RAIN, is
excluded because its host chain could not be verified.

Two structural lessons stand out, and both reinforce the Bitcoin and
Ethereum analysis above.

\begin{itemize}
\tightlist
\item
  \textbf{The exposure model matters as much as the scheme.} This is the
  same asymmetry that separates Bitcoin from Ethereum (§5.3), now
  visible across the market. UTXO
  chains\footnote{UTXO (unspent transaction output) is Bitcoin's accounting model: coins live in discrete, individually locked outputs rather than in a running account balance. With fresh, single-use addresses it keeps the public key hidden until a coin is spent.}
  with fresh, single-use, hashed addresses (Bitcoin, Cardano) and
  stealth or shielded designs (Monero, shielded Zcash) hide the public
  key until a coin is spent, so funds that have never moved are
  protected even against a CRQC; heavy address reuse, as on Dogecoin,
  erodes that protection and is part of why it ranks last.
  \textbf{Account-model chains, including Ethereum, BNB, TRON, XRP,
  Solana, and every token that lives on them, expose the key on the
  first outgoing transaction}, so nearly every active balance is
  permanently exposed. This is why the account-model chains face a more
  urgent ``funds-at-rest'' problem than dormant Bitcoin.
\item
  \textbf{The threat is uniform; the readiness is not.} Grover against
  the hash functions is minor on every chain, and BLS12-381 is a second,
  less-discussed Shor target wherever it secures consensus (Ethereum,
  BNB). The variation in the ranking is almost entirely about migration
  progress, the same governance problem that §7 finds binding for
  Bitcoin and Ethereum.
\end{itemize}

\begin{figure}
\centering
\includegraphics[width=5.5in,height=\textheight,keepaspectratio,alt={Figure 5. Top-20 quantum-readiness ratings; none is fully post-quantum (RAIN excluded; its host chain could not be verified).}]{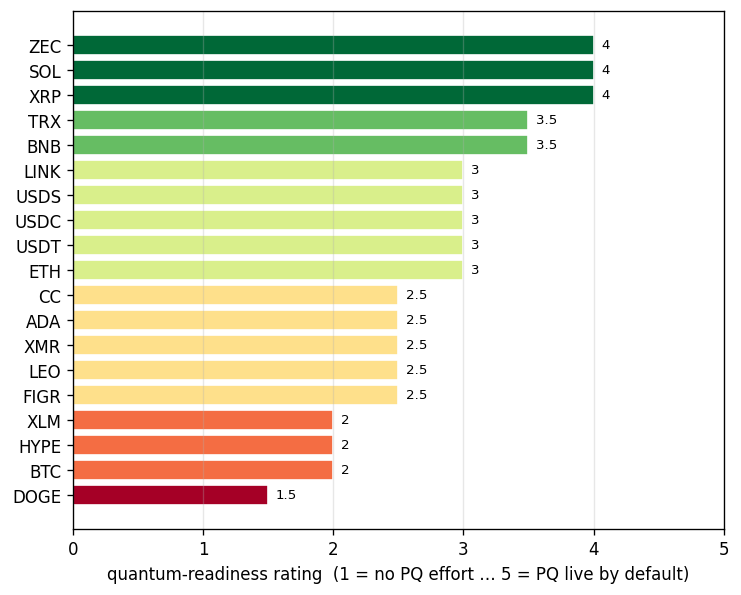}
\caption{\textbf{Figure 5.} Top-20 quantum-readiness ratings; none is
fully post-quantum (RAIN excluded; its host chain could not be
verified).}
\end{figure}

The chains furthest along sit outside the top 20. The \textbf{Quantum
Resistant Ledger} has used hash-based XMSS as its sole signing scheme
since 2018, making it post-quantum by
design;\endnote{``What Is the QRL?'', Quantum Resistant Ledger documentation (2026).
\url{https://docs.theqrl.org/what-is-qrl/}} \textbf{Algorand} has run
Falcon-based state proofs since 2022 and executed a Falcon-signed
mainnet transaction in 2025, though its consensus layer still relies on
classical
signatures.\endnote{``Post-quantum technology,'' Algorand (2022--2026).
\url{https://algorand.co/technology/post-quantum}} Their existence shows
the engineering is solved; the laggards' problem is deployment and
governance, not cryptography. (Two caveats: market-cap rank near the
bottom of the top 20 shifts daily, and most of the migration programs
above are on test or research networks, not yet the production default.)

\begin{quote}
\textbf{General Summary.} Across the 20 largest coins, the picture is
the same as for Bitcoin and Ethereum: the signatures they use today
could all be broken by a future quantum computer, and none has finished
upgrading. A few (XRP, Solana, Zcash) are noticeably ahead with working
test versions; Bitcoin and Dogecoin are among the furthest behind. What
protects a coin most is not which math it uses but whether it hides keys
until you spend, and whether its community has actually started the
upgrade.
\end{quote}

\begin{center}\rule{0.5\linewidth}{0.5pt}\end{center}

\subsection{7. Mitigation and migration}\label{mitigation-and-migration}

\subsubsection{7.1 The tools exist}\label{the-tools-exist}

NIST finalized post-quantum signature
standards\footnote{Post-quantum signatures rely on math problems that quantum computers are not known to solve, either lattice problems (used by ML-DSA and Falcon) or hash functions (used by SLH-DSA, XMSS and LMS). NIST, the U.S. National Institute of Standards and Technology, standardized the first set in 2024.}
in \textbf{August 2024}: ML-DSA (lattice-based), SLH-DSA and the
stateful XMSS and LMS family (hash-based, the most conservative option),
with Falcon still to come.\endnotemark[1] The practical obstacle is
\textbf{size}. A Bitcoin Schnorr signature is about 64--72 bytes;
ML-DSA-44 is 2,420 bytes, and SLH-DSA runs up to about 49,856 bytes, a
10$\times$ to 700$\times$ increase (the 700$\times$ being the largest
SLH-DSA variant, not the sizing a chain would actually choose) that
pressures block space during any mass migration. This is why designs
that commit to a hash of the post-quantum key, and the smaller Falcon
(around 10$\times$), are favored.

\subsubsection{7.2 Bitcoin: feasible but
contested}\label{bitcoin-feasible-but-contested}

\textbf{BIP-360}, a new output type carrying hash-based spending paths,
was merged into the Bitcoin Improvement Proposals repository in February
2026,\endnote{BIP-360, ``Pay-to-Merkle-Root (P2MR)'' (Hunter Beast, Ethan Heilman,
Isabel Foxen Duke), Bitcoin Improvement Proposals, Draft, revised
February 2026.
\url{https://github.com/bitcoin/bips/blob/master/bip-0360.mediawiki}}
though a merge is documentation, not activation. The harder question is
the exposed and dormant coins. \textbf{BIP-361} proposes sunsetting the
legacy signatures, which effectively \emph{freezes} quantum-vulnerable
coins, and it met overwhelmingly negative
reception.\endnote{BIP-361, ``Post Quantum Migration and Legacy Signature Sunset'' (Jameson
Lopp et al., 2026); see also Jameson Lopp, ``Against Allowing Quantum
Recovery of Bitcoin'' (2025).
\url{https://blog.lopp.net/against-quantum-recovery-of-bitcoin/}} This
is the governance crux: a freeze touches the spendability of
otherwise-valid coins, including the roughly 1 million presumed-lost
Satoshi coins, which no prior Bitcoin change has ever done.

\subsubsection{7.3 Ethereum: a clearer
path}\label{ethereum-a-clearer-path}

Account abstraction enables per-account migration with no network-wide
``flag day,'' and it is moving: \textbf{EIP-8141} (``Frame
Transactions,'' drafted January 2026 and under consideration for the
Hegotá fork in late 2026) decouples transaction authorization from
ECDSA, so an account could adopt a post-quantum verifier without
changing its address.\endnotemark[19] Vitalik Buterin's
\textbf{quantum-recovery hard-fork} sketch lets legitimate owners prove
knowledge of their hash-derived seed using a STARK and migrate, while
seedless forgers are frozen
out.\endnote{Vitalik Buterin, ``How to hard-fork to save most users' funds in a
quantum emergency,'' Ethereum Research (March 2024).
\url{https://ethresear.ch/t/how-to-hard-fork-to-save-most-users-funds-in-a-quantum-emergency/18901}}
The Ethereum Foundation's ``Lean Ethereum'' roadmap targets hash-based
validator signatures on a roughly four-year horizon.\endnotemark[18]

\subsubsection{7.4 The race (Mosca's
inequality)}\label{the-race-moscas-inequality}

The decision rule is
Mosca's:\endnote{Michele Mosca, ``Cybersecurity in an Era with Quantum Computers: Will We
Be Ready?'', \emph{IEEE Security \& Privacy} 16(5) (2018); IACR
Cryptology ePrint 2015/1075. \url{https://eprint.iacr.org/2015/1075}}
assets are at risk whenever the time to start migrating plus the time to
migrate exceeds the time until a CRQC arrives. Our migration-race model
sweeps these variables.

\begin{figure}
\centering
\includegraphics[width=5.5in,height=\textheight,keepaspectratio,alt={Figure 4. When migration finishes versus when a quantum computer might arrive, across start-time and adoption scenarios.}]{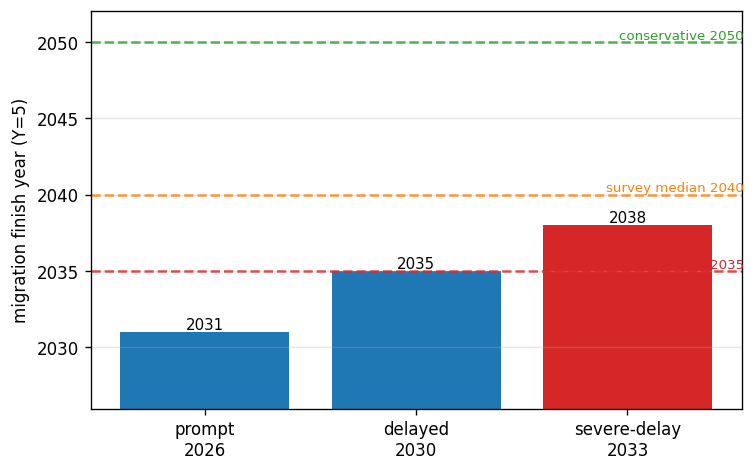}
\caption{\textbf{Figure 4.} When migration finishes versus when a
quantum computer might arrive, across start-time and adoption
scenarios.}
\end{figure}

With a \textbf{prompt start in 2026}, migration finishes around
2029--2031 and \textbf{beats even an optimistic 2035 CRQC with years to
spare}, across every adoption scenario we tried. The \emph{only} at-risk
case in the entire sweep is a severely delayed start (around 2033, that
is, governance paralysis) running against an early machine, and a finer
grid would surface other late-start, early-machine pairings too, so the
conclusion is qualitative. The binding constraint is therefore
\textbf{governance, not throughput or technology}, and the roughly 2.3
million dormant coins are lost regardless (§4.2). The mitigability and
residual-concentration findings hold: this is substantially fixable, but
the window is finite.

\begin{quote}
\textbf{General Summary.} The cryptographic fixes already exist,
standardized in 2024, and both networks have migration plans. The hard
part is not the math, it is coordination: getting a decentralized
community to agree, especially Bitcoin's, where the choice between
protecting lost coins and freezing them is contentious. If migration
starts soon it finishes with years to spare; the danger is delay, not
difficulty.
\end{quote}

\begin{center}\rule{0.5\linewidth}{0.5pt}\end{center}

\subsection{8. The strongest case for
alarm}\label{the-strongest-case-for-alarm}

The case for greater alarm deserves a fair hearing. Its strongest form,
and the reasons it does not overturn the thesis:

\textbf{The case.}

\begin{enumerate}
\def\labelenumi{\arabic{enumi}.}
\tightlist
\item
  \textbf{The timeline is collapsing faster than anyone predicted.} The
  2025 result cut RSA-2048 about twentyfold, from the 2019 baseline of
  20 million
  qubits\endnote{Craig Gidney and Martin Ekerå, ``How to factor 2048 bit RSA integers in
  8 hours using 20 million noisy qubits,'' \emph{Quantum} 5 (2021),
  arXiv:1905.09749. \url{https://arxiv.org/abs/1905.09749}} to under 1
  million,\endnotemark[2] with no hardware change at all, pure
  algorithmic progress that can recur without warning. The whitepaper
  that put Bitcoin's curve at under 500,000 physical qubits was written
  by Babbush, Gidney, Drake, and Boneh, not by fringe
  voices.\endnotemark[3] Expert ten-year probabilities are the highest
  ever recorded.\endnotemark[7]
\item
  \textbf{``Harvest now, decrypt later'' is already
  live.}\footnote{Harvest now, decrypt later: capturing data that is exposed today so it can be broken later, once a quantum computer exists. Public keys already on the blockchain can be collected now and attacked whenever a CRQC arrives.}
  Every exposed public key (every used Ethereum account, every reused
  Bitcoin address, all Taproot and P2PK coins) is a standing target a
  future machine can cash in. A fork cannot retroactively protect them,
  and the Federal Reserve's 2025 analysis concludes that prior on-chain
  data ``remains
  vulnerable.''\endnote{Jillian Mascelli and Megan Rodden, ``\,`Harvest Now, Decrypt Later':
  Examining Post-Quantum Cryptography and the Data Privacy Risks for
  Distributed Ledger Networks,'' Federal Reserve Finance and Economics
  Discussion Series 2025-093 (September 2025).
  \url{https://www.federalreserve.gov/econres/feds/files/2025093pap.pdf}}
\item
  \textbf{Governance may not deliver in time.} Bitcoin's overwhelmingly
  negative response to a freeze proposal suggests the political system
  may \emph{not} execute a timely migration, and the migration-race
  model shows that delay is the one path to disaster.
\item
  \textbf{The numbers are large.} Millions of BTC and tens of millions
  of ETH, hundreds of billions of dollars, sit in exposed accounts.
\end{enumerate}

\textbf{Why it does not overturn the thesis.}

\begin{enumerate}
\def\labelenumi{\arabic{enumi}.}
\tightlist
\item
  \textbf{Algorithmic re-estimation is not hardware.} Every 2026
  reduction shrinks the \emph{requirement}; none builds the machine. The
  gap stays large, about 400--500$\times$ in physical qubits even
  against the most aggressive estimate and far more against conservative
  ones, and no fault-tolerant system exists at scale. The timing is
  contested, not settled: the combined forecast puts the chance of a
  CRQC by 2035 at roughly one-in-six, and below one-in-four even on its
  survey-weighted end (§3.2). That uncertainty cuts against confident
  alarm as much as against complacency.
\item
  \textbf{The unfixable core is bounded and known.} The set that can be
  harvested and never repaired is the roughly 2.3 million dormant BTC
  plus other lost-key coins, a real and permanent loss but a
  \emph{bounded and characterizable} one, not an open-ended systemic
  collapse. Everything else is either protected-until-spent or
  migratable.
\item
  \textbf{Governance is solvable, and the clock is still generous.} A
  prompt start wins by years; the remedy is simply to \emph{start}, and
  Ethereum already has. The risk is delay, which is actionable today.
\item
  \textbf{Size is not the same as fragility.} Large exposed balances at
  \emph{active} accounts are exactly the ones that can be swept to
  safety on warning. And the networks' core security functions,
  proof-of-work and hashing, are untouched (§2.2).
\end{enumerate}

The case wins two points. ``The threat is far off'' is no longer true,
and governance delay is the live danger. The thesis narrows accordingly,
to bounded and mitigable \textbf{but compressing, and contingent on
acting now}.

\begin{quote}
\textbf{General Summary.} The strongest argument against this paper is
that the timeline is collapsing and that lost coins can never be
protected. Both points are real, and they rule out any claim that the
threat is comfortably far away. But they do not overturn the core
finding: no machine exists yet, mining is safe, and the great majority
of coins can be moved to safety in time, provided the work starts soon.
\end{quote}

\begin{center}\rule{0.5\linewidth}{0.5pt}\end{center}

\subsection{9. Residual risk: the irreducible
core}\label{residual-risk-the-irreducible-core}

Strip away everything that is protected (the 65--70\% of Bitcoin in
fresh addresses, the ETH held in contracts, the slow and buffered
consensus layer) and everything migratable (the roughly 3.7 million
movable BTC, the 50--65\% of ETH in active accounts that can adopt new
signatures), and what remains is small and well-defined:

\begin{itemize}
\tightlist
\item
  the roughly \textbf{2.3 million BTC} that is dormant or lost and can
  never be voluntarily migrated, the hard floor (Ethereum's
  dormant-and-exposed counterpart is far smaller, on the order of 0.1\%
  of supply\endnotemark[18]); and
\item
  \textbf{the transition window}, the period during migration when coins
  are briefly exposed to a fast-clock mempool snipe.
\end{itemize}

That is the shape of the danger. It is not a sudden break of two
trillion-dollar networks, but a bounded, foreseeable loss concentrated
in coins nobody can move, plus a manageable transition risk. It is
serious, since the eventual theft of Satoshi-era coins would be a
confidence shock, but it is neither systemic nor unmanageable.

\begin{quote}
\textbf{General Summary.} Once you set aside everything that is safe or
fixable, what is left is small and predictable: a few million lost
Bitcoin that nobody can rescue, plus a brief risky stretch during the
upgrade itself. That would make headlines if it happened, but it is a
long way from the collapse of the two networks.
\end{quote}

\begin{center}\rule{0.5\linewidth}{0.5pt}\end{center}

\subsection{10. Conclusion and
recommendations}\label{conclusion-and-recommendations}

Quantum computing is a real threat to Bitcoin and Ethereum, aimed
squarely at their \textbf{signature} schemes, on a timeline that
compressed meaningfully in 2025--26 but still places a code-breaking
machine years away and absent today: a wide, bimodal forecast with a
real near-term tail and a center of mass in the 2040s. The exposure is
\textbf{bounded and mostly migratable}, the proof-of-work and hashing
layers are \textbf{safe}, and the binding constraint on a good outcome
is \textbf{governance speed}, not technology. The same pattern holds
across the wider market: the cryptography is uniformly vulnerable and
uniformly fixable, and what separates the chains is how soon they act.

\textbf{Recommendations.}

\begin{itemize}
\tightlist
\item
  \textbf{Holders:} never reuse addresses; keep funds in fresh
  hash-protected outputs on Bitcoin and minimize exposed-account
  balances on Ethereum; for large sums, wait for confirmations; and plan
  to migrate to post-quantum-secured outputs as they ship.
\item
  \textbf{Protocols:} advance and \emph{activate} the post-quantum
  migration paths now (BIP-360 and its successors on Bitcoin, account
  abstraction and the hash-based-signature roadmap on Ethereum); settle
  the dormant-coin governance question now, on a considered timeline,
  rather than under emergency pressure; and treat the migration window
  itself as a security-critical event.
\item
  \textbf{Everyone:} track the timeline. A single algorithmic result
  moved it roughly twentyfold in 2025. The posture is prepared urgency,
  not alarm.
\end{itemize}

\begin{quote}
\textbf{General Summary.} The bottom line is a manageable,
signature-focused threat on a finite clock. Individuals should use fresh
addresses and plan to upgrade; the networks should switch on their
quantum-proof upgrades now rather than wait for an emergency. Prepared,
not panicked.
\end{quote}

\begin{center}\rule{0.5\linewidth}{0.5pt}\end{center}

\subsection{Disclosures and
availability}\label{disclosures-and-availability}

\textbf{Affiliations.} Iosif M. Gershteyn is the chief executive of
ImmuVia and chairs the Ajax Biomedical Foundation. Jacob A. Alber is the
founder of Sataresse AI.

\textbf{Competing interests.} Both authors have at various times held
cryptocurrencies and related instruments among those discussed in this
paper, including Bitcoin and Ether. Iosif M. Gershteyn is additionally
the owner of subnet 103 (SN103) on the Bittensor network; Bittensor's
native token (TAO) is outside the survey of §6 but, like every asset
analyzed here, relies on elliptic-curve signatures and faces the same
class of quantum risk.

\textbf{Use of AI tools.} AI tools were used during the research and
drafting of this paper, including for literature search, modeling
support, and editorial passes. The authors reviewed the claims,
citations, and computations, and take full responsibility for the
content.

\textbf{Code and data availability.} The model code, figures, and full
numerical results are available at
\url{https://github.com/imgcode/quantum-horizon}.

\begin{center}\rule{0.5\linewidth}{0.5pt}\end{center}

\subsection{Appendix A. Methodology and
models}\label{appendix-a.-methodology-and-models}

The quantitative claims rest on a suite of small, reproducible models,
built so that independent ones could test each other, with one rule: a
model is credited with confirming another only insofar as it shares no
inputs, assumptions, or method with it. The timeline is estimated two
ways, a bottom-up physical-qubit scaling model and an expert-survey
model, and then combined into the single systemic forecast of §3.2
(which blends rather than cross-validates, and is labeled as a combined
estimate, not a confirmation). Two exposure models quantify the at-risk
supply on Bitcoin and Ethereum; a mining-competitiveness model settles
the proof-of-work question; and a mempool-race model and a
migration-race model are downstream of the timeline, consumers of its
output and never validators of it. Calibration passes against the 2017
Aggarwal benchmark (mining),\endnotemark[9] the published physical-qubit
requirements (timeline),\endnotemark[3] and the literature's 41\%
sniping figure (mempool race). The market survey of §6 is a sourced
field assessment, not a model. The model code and full results are
available in the supplementary repository (see Disclosures and
availability).

\subsection{Appendix B. Limitations and open
questions}\label{appendix-b.-limitations-and-open-questions}

\begin{itemize}
\tightlist
\item
  The bottom-up timeline model treats physical-qubit scaling as the
  binding constraint, and the fault-tolerance lag added in the systemic
  forecast is a modeling assumption rather than a measured quantity; it
  shifts the later (physics) mode but barely changes the near-term
  probability, which is set by the survey weight (§3.2). The combined
  distribution is bimodal, and its median is reported only as the
  midpoint between two disagreeing estimators, not as a best estimate.
\item
  Ethereum's at-rest percentage rests on a verified 2021 on-chain
  measurement,\endnotemark[5] hardened for 2026 to 50--65\% (center
  55--60\%) by two independent reconstructions that converge. A June
  2026 search found no public re-measurement anywhere (every 2025--26
  figure in circulation traces back to that same scan), so a fresh
  recount on current ledger data remains the one outstanding data task.
\item
  BLS12-381 has no direct published resource estimate, so we interpolate
  (about 1,950 logical qubits) and flag it as such.
\item
  The §6 market survey reflects a mid-2026 snapshot; market-cap ranks
  near the bottom of the list shift daily, one entry could not be
  verified and is excluded, and most cited migration programs are on
  test or research networks rather than in production.
\end{itemize}

\begin{center}\rule{0.5\linewidth}{0.5pt}\end{center}

\clearpage
\theendnotes

\end{document}